\def\year{2017}\relax
\documentclass[letterpaper]{article}

\usepackage{aaai17}
\usepackage{times}
\usepackage{helvet}
\usepackage{courier}
\usepackage{graphicx}
\usepackage{url}
\usepackage{multirow}
\usepackage{color}
\usepackage{subfig}
\usepackage{amsmath}
\usepackage[T1]{fontenc}

\newcommand{\superscript}[1]{\ensuremath{^{#1}}}

\def\iit{\superscript{*}}
\def\mpi{\superscript{\#}}
\def\ufmg{\superscript{o}}

\begin{document}

\title{Who Makes Trends?\\Understanding Demographic Biases in Crowdsourced Recommendations}

\author{
  Abhijnan Chakraborty\iit\mpi,~
  Johnnatan Messias\ufmg\mpi,~
  Fabricio Benevenuto\ufmg,
  \AND
  Saptarshi Ghosh\iit,~~
  Niloy Ganguly\iit,~~
  Krishna P. Gummadi\mpi \\
  ~\\
  {\bf {\mpi}Max Planck Institute for Software Systems, Germany} \\
  {\bf {\iit}Indian Institute of Technology Kharagpur, India} \\
  {\bf {\ufmg}Universidade Federal de Minas Gerais, Brazil} \\
}

\maketitle

\begin{abstract}
\noindent Users of social media sites like Facebook and Twitter rely
on crowdsourced content recommendation systems (e.g., Trending Topics)
to retrieve important and useful information. Contents selected for
recommendation indirectly give the initial users who promoted (by
liking or posting) the content an opportunity to propagate their
messages to a wider audience. Hence, it is important to understand the
demographics of people who make a content worthy of recommendation,
and explore whether they are representative of the media site's
overall population. In this work, using extensive data collected from
Twitter, we make the first attempt to quantify and explore the
demographic biases in the crowdsourced recommendations. Our analysis,
focusing on the selection of trending topics, finds that a large
fraction of trends are promoted by crowds whose demographics are
significantly different from the overall Twitter population. More
worryingly, we find that certain demographic groups are systematically
under-represented among the promoters of the trending topics. To make
the demographic biases in Twitter trends more transparent, we
developed and deployed a Web-based service `Who-Makes-Trends' at 
{\tt twitter-app.mpi-sws.org/who-makes-trends}.
\end{abstract}

\section{Introduction}
\label{sec:intro}
\noindent Social media sites like Facebook and Twitter have emerged as
popular destinations for users to get real-time news about the world
around them. In these sites, users are increasingly relying on
crowdsourced recommendations called {\bf Trending Topics} to find
important events and breaking news stories.  Typically topics,
including key-phrases and keywords (e.g., hashtags), are recommended
as trending when they exhibit a sharp spike in their popularity, i.e.,
their usage by the crowds suddenly jump at a particular
time~\cite{twitter_trends}.  Once a topic is selected as trending, it
gets prominently displayed on the social media homepages, thus
reaching a large user population.  Additionally, traditional news
organizations often pick the trending topics and cover them in their
news stories (a practice termed as {\it Hashtag
  journalism}~\cite{Friedman2014Hashtag}), further amplifying their
reach.  Recognizing their importance, researchers have started arguing
whether trending topics have become a part of our societal
culture~\cite{gillespie2016trendingistrending}.

A large number of prior works on trending topics have focused on {\bf
  what} the trends are (e.g., classifying the trends into different
categories~\cite{naaman2011hip}), or {\bf how} the trends are selected
(e.g., proposing new algorithms to identify trends from the content
stream~\cite{benhardus2013streaming}).  Complementary to the earlier
works, our focus in this paper is on the users {\bf who} make
different topics worthy of being recommended as
trending. Specifically, we attempt to analyze the {\it demographics}
of crowds {\it promoting} different topics on the social media
sites. By {\it promoters} of a topic, we refer to the users who posted
on the topic {\it before} it became trending, thereby contributing to
the topic's selection as a trend. 

In this paper, our focus is on the {\it biases in the demographics} of
the promoters of different trends, i.e., we investigate whether the
distribution of trend promoters across different socially salient
groups are representative of the media site's overall user
population. As users belonging to different demographic groups (such
as middle-aged white men, young asian women, adolescent black men)
might be interested in posting about different topics, the
demographics of a topic's promoters can be quite different from the
site's user population. Our goal here is to study the demographic
biases of trends, i.e., we quantify and analyze the divergence between
the demographics of the promoters of trends and the site's overall population.

Towards this end, we gathered extensive data from
the popular social media site Twitter over a period of 3 months from
July, 2016 to September, 2016. Our data included over five thousand
trending topics recommended to Twitter users in the United States, and
millions of users posting on the topics, both before and after the
topics became trending. We inferred three demographic attributes for
these Twitter users namely, their gender, race, and age.
We performed a detailed analysis on these users, and our
analysis offers several interesting insights:

{\bf (a)} We find that a large fraction of trending topics are
promoted by crowds whose demographics are {\it significantly}
different from Twitter's overall user population.

{\bf (b)} We find clear evidence of 
under-representation of certain demographic groups (female, black, mid-aged)
among the promoters of the trending topics, with 
mid-aged-black-females being the most under-represented group.

{\bf (c)} We discover that once a topic becomes trending, it is {\it
  adopted} (i.e., posted) by users whose demographics are less
divergent from the overall Twitter population, compared to the users
who were promoting the topic {\it before} it became trending.  Our
finding alludes to the influence and importance of trending topic
selection on making users aware of specific topics.

{\bf (d)} We observe that topics promoted predominantly by a single
demographic group tend to be of niche interest to that particular
group.  However, during events of wider interest (e.g., national
elections, police shootings), 
the topics promoted by different demographic groups tend to reflect
their diverse perspectives, which could help understand the different
facets of public opinion.

Our findings make the case for making the demographic biases of
Twitter trend recommendations transparent. Therefore, 
we developed and deployed a system `Who-Makes-Trends'\footnote{\tt 
twitter-app.mpi-sws.org/who-makes-trends}, where for any trend in
the US, one can check the demographics of the promoters of that trend.
We further note that our analysis framework and findings about demographic
biases can be extended to other social media algorithms such as social
search or social influence estimation.

More broadly, our work offers a new perspective to the growing debate
about fairness, bias, and transparency of decisions taken by
algorithms operating over big crowdsourced data~\cite{zafar2015learning,chakraborty2016dissemination,kroll2015accountable}.
Similar to the work by Kulshrestha {\it et al.}~\cite{Kulshrestha2017}, 
it highlights the need to understand the characteristics and biases in the 
inputs to the algorithms (e.g., the group of users promoting specific trending topics), in addition to
studying the algorithms (e.g., selection of trending topics), and
their outputs. 

\section{Related Work}
\label{sec:related}
\noindent
We briefly survey related efforts in the following three axes. First, we discuss efforts that explore social media demographics. Then, we point to the studies focusing on Twitter trends. Finally, we survey a few approaches for %that attempt to
providing transparency to algorithms and systems.

\vspace{1mm}
\noindent \textbf{Social Media Demographics: }
Recently, there has been a growing interest for demographic aspects of social media data.
Mislove {\it et al}~\cite{mislove2011understanding} provided one of the first studies 
in this space, by looking at the gender and racial demographics
of Twitter users, and analyzing how the demographics vary across different US states.
Pew research~\cite{pew_demography} conducted user surveys to understand the demographics
of users in different social media platforms. There have also been past attempts to understand
the use of social media among particular demographic groups. For example, Madden {\it et al.}~\cite{madden2013teens}
explored how teenagers use different social media sites. Gilbert {\it et al.}~\cite{gilbert2008network}
analyzed social media use among rural users.
In another research direction, many efforts attempt to quantify inequalities in social media systems, including Wikipedia~\cite{wagner2016women}, 
Pinterest~\cite{Gilbert:2013:INT:2470654.2481336}, and Twitter~\cite{nilizadeh2016twitter}.
However, to the best of our knowledge, we are not aware of any effort that approached the demographics
behind crowdsourced recommendations deployed in social media sites.
Thus, our endeavor is complementary to the above mentioned approaches.

\vspace{1mm}
\noindent \textbf{Twitter Trends Analysis:}
Many prior works have focused on 
Twitter Trending Topics. 
For example, Zubiaga \textit{et al.} presented an approach to automatically categorize trending topics~\cite{zubiaga2011classifying}.
Lee \textit{et al.} characterized how spammers can exploit trending topics~\cite{lee2012detecting}.
Benhardus \textit{et al.}~\cite{benhardus2013streaming} proposed alternative algorithms for detecting trending topics.
Our prior work~\cite{chakraborty2015can} identified temporal coverage biases in the selection of trending stories.

Most of the existing efforts on this space have focused on the {\it outputs} (results) from the trending topics selection algorithms.
Thus, our work offers a novel and complementary angle as we focus on the {\it input} 
to the Twitter trend selection algorithms.
To the best of our knowledge, there has been no prior attempt to 
analyze the demographic distribution of the crowd who make a particular topic trending.

\vspace{1mm}
\noindent \textbf{Algorithmic and Data Transparency: }
Increasingly, researchers and governments are recognizing the importance of making algorithms transparent.
The White House recently released a report that concludes that the practitioners must ensure that AI-enabled
systems are open, transparent, and understandable~\cite{white_house_transparency}.
Indeed, the controversy around Facebook using human editors
on their trending topics teaches a lesson about the
importance of transparency. On one hand, when humans were editing trending topics,
they were accused to select and filter content~\cite{facebook_bias}.
On the other hand, when humans were removed from the process,
Facebook was accused of featuring fake news as trending~\cite{Ohlheiser2016}.

Our present effort contributes to make the demographic biases of Twitter 
trend recommendations transparent, and we hope that the methodology
to compute demographic distribution of users can be leveraged to make
other crowdsourced systems (e.g., social search~\cite{Kulshrestha2017}) 
transparent as well. More importantly, for algorithms that operate on 
large-scale crowdsourced data, we show that along
with making the outputs of the algorithms (and the algorithm itself) transparent,
it is also important to understand the non-uniformities in the inputs to the algorithms.

\section{Methodology and Dataset}
\label{sec:method}
\noindent
In this section, we first describe the dataset gathered, then the method to
infer demographic information of individual Twitter users.

\vspace{1mm}
\noindent \textbf{Twitter dataset gathered} \\
\noindent For this work, we gathered the $1\%$ random sample of all tweets, 
through the Twitter Streaming API\footnote{\it dev.twitter.com/streaming/public}, 
along a $3$-month period, from July to September, 2016. Simultaneously and along the same period, 
by querying the Twitter REST API\footnote{\it dev.twitter.com/rest/reference/get/trends/place} 
at every $5$-minutes, we collected all topics which became trending in US.

In total, we collected more than $340$ million tweets posted by around $50$ million users.
During these three months, $11,797$ topics became trending, out of which
$5,810$ ($49.25\%$) trends were {\it hashtags} and the rest were multi-word phrases.
For simplicity, we restrict our focus on trending hashtags, and leave the analysis of trending phrases as  future work.

\vspace{1mm}
\noindent \textbf{Inferring demographics of Twitter users} \\
\noindent While conducting traditional user interviews, 
social survey agencies like Pew Research 
typically ask the respondents different aspects of user
demographics, including the gender, race, age group, location,
educational qualification, or the annual income of the users.
To collect Twitter users' demographic information at scale,
we can only use publicly available information about a user,
such as her name, profile description, location, profile image,
and the tweets she posted. Due to this limitation, in this work, we consider three aspects of
user demographics -- gender, race, and age group, and we restrict our analysis on users
whose location could be identified as within US.

Past works have attempted to infer a particular user's gender and race
from her name~\cite{blevins2015jane,mislove2011understanding,liu2013s}, or
the age from Twitter profile description
(by searching for patterns like {\it `21 yr old'} or {\it `born in 1989'})~\cite{sloan2015tweets}.
However, Liu {\it et al.}~\cite{liu2013s} reported that $66\%$ users in their
dataset did not have a proper name and hence their gender could not be inferred.
Similarly, to infer the age from the profile descriptions,
we could find age related patterns for only $0.2\%$ of the users in our dataset.
To circumvent the difficulties in inferring the
demographic information from users' profile names and descriptions, we
use the profile pictures of the users to get their demographic information.
Specifically, we use {\it Face++} ({\tt faceplusplus.com}), 
a face recognition platform based on deep learning~\cite{yin2015learning},
to extract the gender, race, and age information from the recognized
faces in the profile images of all US based users in our dataset.

We observed four issues with using the profile images for inferring
demographics: 
(i)~some users may have Twitter's default profile picture, while
others have customized profile images, 
(ii)~a profile image may not have any recognizable face, 
(iii)~multiple faces can be present in
an image (e.g., group photo), and
(iv)~some users may  change their profile pictures between the time the tweets are 
collected and the time at which the demographic inference is attempted. 
To address the first issue, we check the URLs of the profile images and discard the users having default 
profile pictures. For issues (ii) and (iii), we check the output of Face++, and users whose
profile images contain zero or more than one faces are discarded.
Finally, when users change their picture, the corresponding URL changes as well, 
making it impossible for us to gather demographic information for such users;
hence we ignore such users. 
In our dataset, we have around $4$ million US based users with valid profile image URLs. 
After performing the filtering steps discussed above, we consider the demographic
information, as returned by Face++, for around $1.7$ million users. 

\vspace{1mm}
\noindent \textbf{Possible values of the demographic attributes}\\
\noindent
Face++ returns the values \textbf{\textit{Male}} or \textbf{\textit{Female}} for the gender,
\textbf{\textit{White, Black}}, or \textbf{\textit{Asian}} for the race,
and a numerical value corresponding to the estimated age of the recognized face in the profile image.

In this work, we use the values of gender and race as returned by Face++.
To form the age groups, we bucketize the age values according to the seminal
work by Erikson~\cite{erikson1994identity},
where he proposed eight stages of psychosocial development in human life-cycle.
Discarding the first four childhood stages, we use the remaining
four stages of adulthood as the age groups in this work.
Specifically, we use the following four age groups:
(i) \textbf{\textit{Adolescent}} (13 -- 19 years),
(ii) \textbf{\textit{Young}} ({\it `Early adulthood'} in Erikson's parlance) (20 -- 39 years),
(iii) \textbf{\textit{Mid-aged}} ({\it `Adulthood'}) (40 -- 64 years), and
(iv) \textbf{\textit{Old}} ({\it `Maturity'}) (65 years and above).

\vspace{1mm}
\noindent \textbf{Evaluating the demographic inference by Face++ }\\
\noindent
Along with the inferred demographic information, Face++ returns confidence levels for inferred gender and race,
and an error range for inferred age. The average confidence levels reported by Face++ for our data
were $95.03\pm0.02\%$ for gender and $85.99\pm0.03\%$ for race inference, respectively.
The average error range reported for age inference was $6.53\pm0.0038$ years.

To further evaluate the effectiveness of the inference made by Face++,
we asked $3$ volunteers to label $100$ randomly selected profile images from our dataset.
We measured the inter-annotator agreement in terms of the {\it Fleiss'} $\kappa$ score.
For gender labeling, $\kappa$ score was $1.0$ denoting complete agreement, $\kappa$ was $0.865$ for race,
and regarding labeling age group, $\kappa$ was $0.58$, implying that inferring the exact age group is tough
even by  humans. It is especially difficult for users having age bordering two age groups.

Comparing the labels made by majority of the human annotators, and the ones inferred by Face++,
we find that the accuracy of gender inference is $88\%$, while the accuracy for race is $79\%$.
If we take the absolute age returned by Face++ (without the error range), age group is correct
in $68\%$ cases. However, if we consider the error range, especially in the border of two age groups,
the accuracy of the age inference shoots up to $83\%$, considering either age group to be correct in such cases.
Separately, there have been some recent attempts to use Face++ for inferring the gender and
age of Twitter users~\cite{Zagheni2014,an2016greysanatomy}.
We note that the results found in our evaluation are comparable with their observations.

\vspace{1mm}
\noindent \textbf{Limitations } \\
\noindent Inferring the age, race, and gender from the profile images
are challenging tasks, and we are limited by the accuracy of Face++ in the inference.
However, as the performance of deep learning systems continue to
improve, the inferred demographic attributes should become more accurate.
The other limitation of using Face++ is that it reports the race of the users but
not the {\it ethnicity} (e.g., {\it Hispanic}). In future work, we aim to explore
alternative approaches to overcome this limitation. 

\section{Analysis}
\label{sec:analysis_promoters}
\noindent
For a trending topic on Twitter, we define the {\bf promoters} of the trend as the users who have tweeted that topic
{\it before} the topic actually became trending.
Therefore, promoters are the users who make different topics worthy of being recommended by Twitter as trending.
In this section, we attempt to analyze the promoters of different trends.

Specifically, our focus is on the demographics of the promoters
of different trending topics on Twitter.
For each topic, we compute the demographics of the promoters
as a {\it vector} where each entry corresponds to the fraction of
the promoters belonging to different demographic groups
(such as middle-aged white men, young asian women, adolescent black men).
Such demographic groups can be considered either using a single demographic
attribute (e.g., only gender, race, {\it or} age) or a combination
of multiple attributes (e.g., gender {\it and} race).
Before analyzing the demographics of the promoters of different trends, we first
look at the demographics of all active Twitter users in US.

\subsection{Demographics of the user population on Twitter}
\noindent
Using the demographic information of around $1.7$ million US based Twitter users, as obtained from Face++,
we compute the overall demographics of such users.
Table~\ref{tab:reference_dist} shows the distribution of gender, race, and age groups among the
US based users.
We can see in Table~\ref{tab:reference_dist} that more women than men, more Whites,
and more young people are present in Twitter. Considering the race and gender together,
 $32.88\%$ of users in Twitter are white men, $35.1\%$ are white women, $6.55\%$ are black men,
 $7.13\%$ are black women, $7.26\%$ are asian men, and $11.13\%$ users are asian women.

To compare the demographics of Twitter users with the
demographics of the {\it offline} population, we collect the demographics of US residents from the
U.S. Census Bureau\footnote{\it census.gov/population/age/data/2012comp.html}
\footnote{\it census.gov/prod/cen2010/briefs/c2010br-02.pdf} 
and present in Table~\ref{tab:reference_dist}.
We see that some demographic groups are present a lot more in Twitter compared to their share of US population.
For example, the presence of Asians in Twitter is about 4 times more than in the overall US population.
Similarly, the adolescent and young people are present significantly more in Twitter. On the contrary, mid-aged and old population
have comparatively much less presence in Twitter. Our findings corroborate with a recent survey on social media
population conducted by Pew Research\footnote{\it pewinternet.org/2016/11/11/social-media-update-2016}.

\begin{figure}[t]
\center{
\includegraphics[width=0.85\columnwidth]{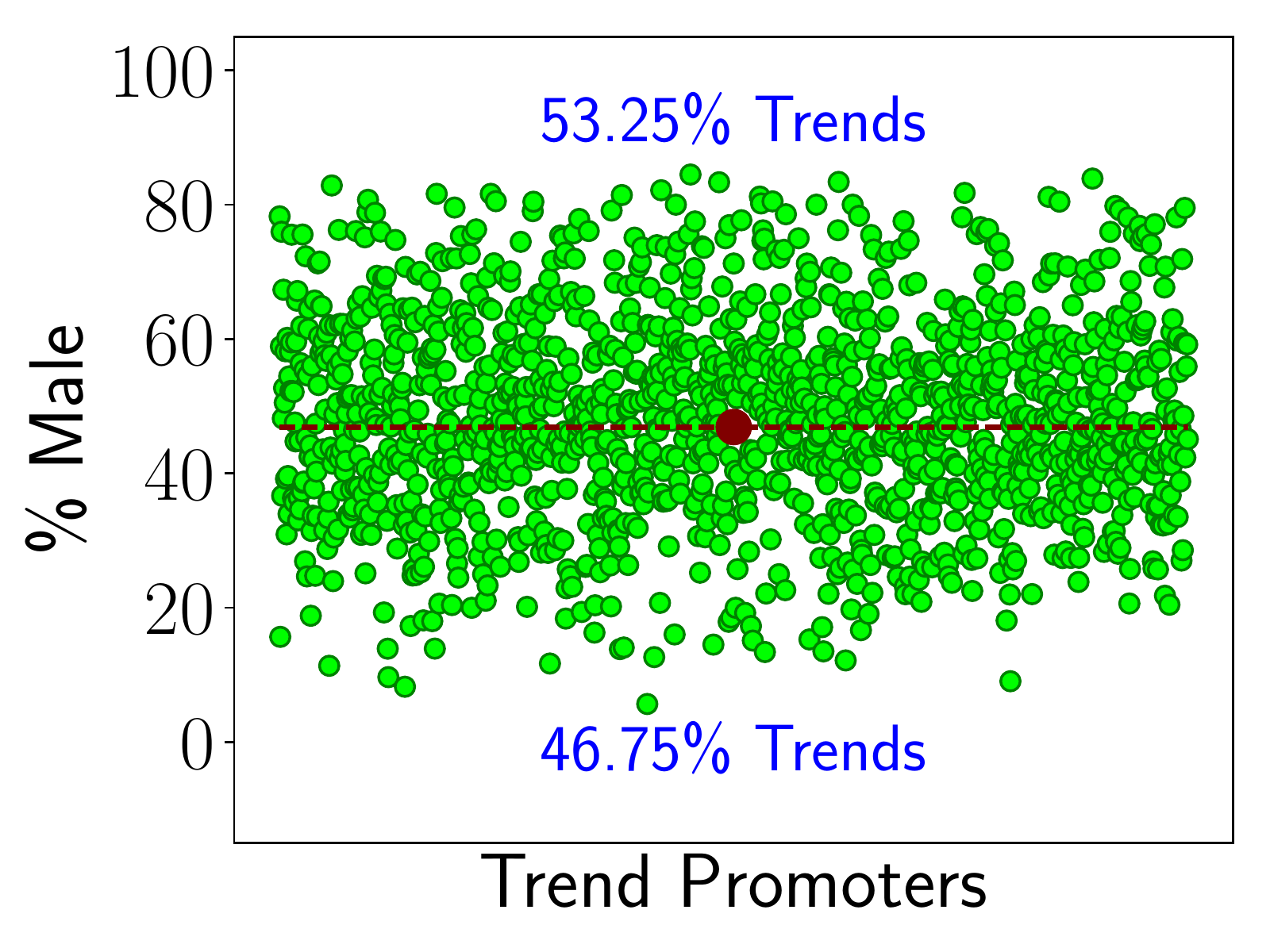}
}
\vspace{-2mm}
\caption{{\bf Gender distribution among the promoters of Twitter trends. Green dots represent the proportion of men among the promoters. The proportion of women can be implicitly derived by taking 100's complement. Red dot represent the proportion of men in overall Twitter population. The red scattered line divides the trends into two halves: (i) upper-half containing the trends ($53.25\%$) where men are represented more among the promoters, and (ii) lower-half containing trends ($46.75\%$) where men are represented less compared to their share in the overall Twitter population.}}
\label{fig:scatter_gender}
\vspace{-5mm}
\end{figure}

\begin{table*}[t]
\center
\small
\begin{tabular}{|p{0.135\textwidth}||p{0.055\textwidth}|p{0.075\textwidth}||p{0.062\textwidth}|p{0.065\textwidth}|p{0.065\textwidth}||p{0.1\textwidth}|p{0.065\textwidth}|p{0.09\textwidth}|p{0.05\textwidth}|}
\hline
\multicolumn{1}{|c||}{{\bf Baseline}} & \multicolumn{2}{c||}{{\bf Gender}} & \multicolumn{3}{c||}{{\bf Race}} & \multicolumn{4}{c|}{{\bf Age Group}}\\
\cline{2-10}
~ & \% Male & \% Female & \% White & \% Black & \% Asian & \% Adolescent & \% Young & \% Mid-aged & \% Old \\
\hline
US Population & 49.2 & 50.8 & 72.4 & 12.6 & 4.8 & 13.6 & 26.7 & 33.2 & 13.5 \\
\hline
Twitter Population  & 46.9 & 53.1 & 67.9 & 13.7 & 18.3 & 29.3 & 61.2 & 9.3 & 0.2 \\
\hline
\end{tabular}
\vspace*{-2mm}
\caption{{\bf Comparing the demographics of the population in US, and the demographics of US based Twitter users, whose tweets were included in the 1\% random sample during July -- September 2016, and whose demographic information could be inferred.}}
\label{tab:reference_dist}
\vspace*{-5mm}
\end{table*}

\begin{figure*}[tb]
\center{
\subfloat[{\bf }]{\includegraphics[width=0.33\textwidth]{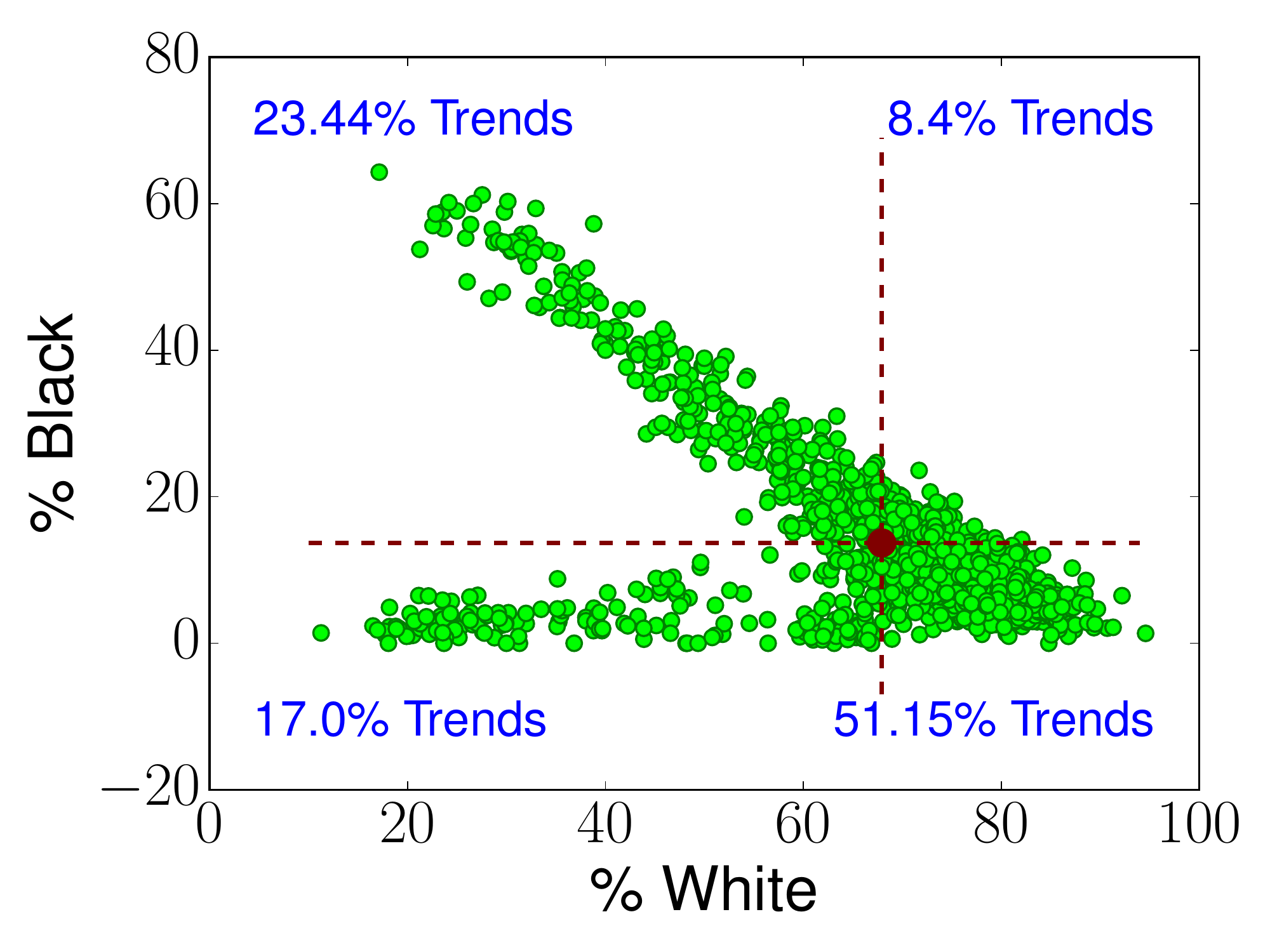}}
\hfil
\subfloat[{\bf }]{\includegraphics[width=0.33\textwidth]{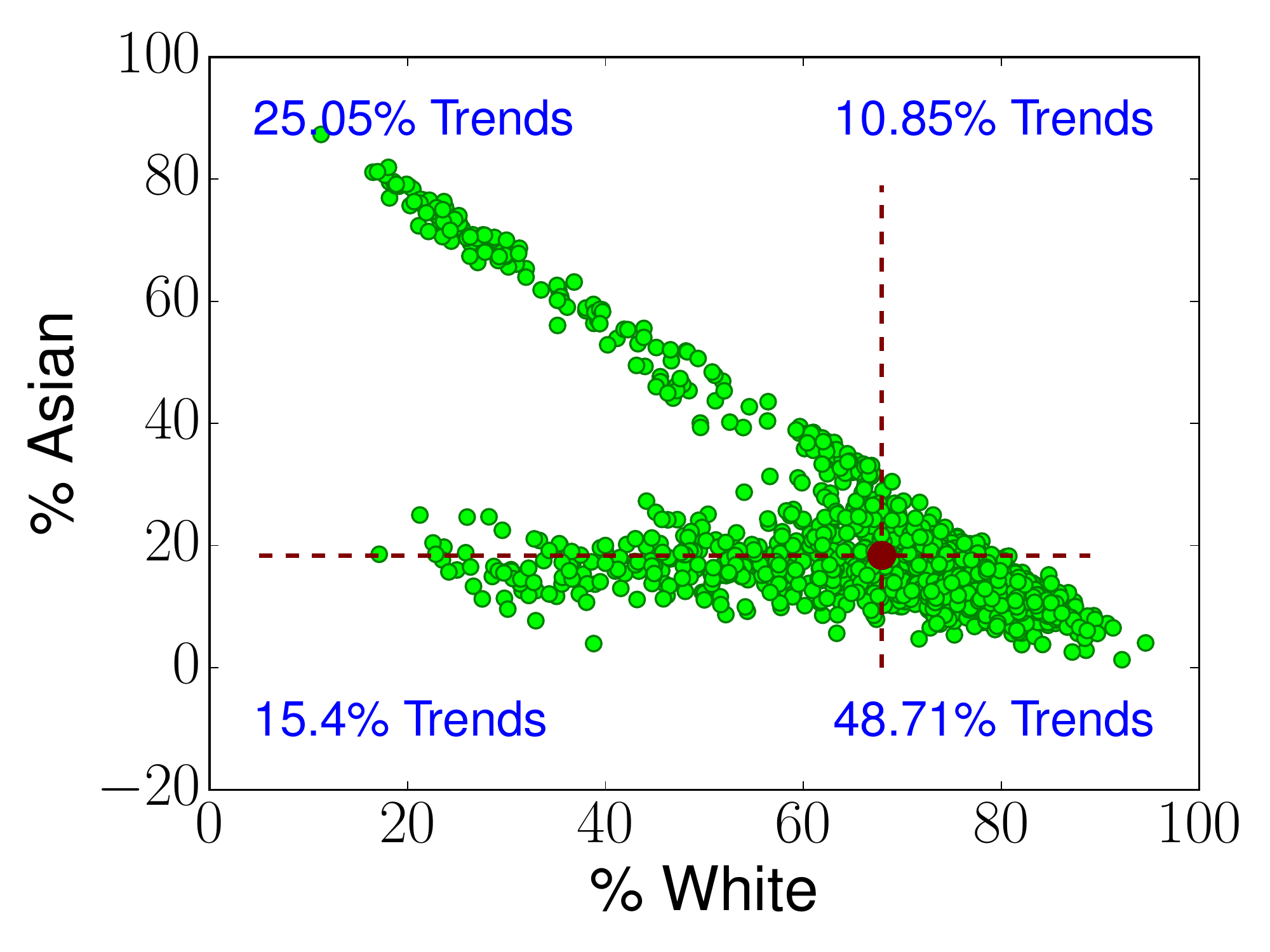}}
\hfil
\subfloat[{\bf }]{\includegraphics[width=0.33\textwidth]{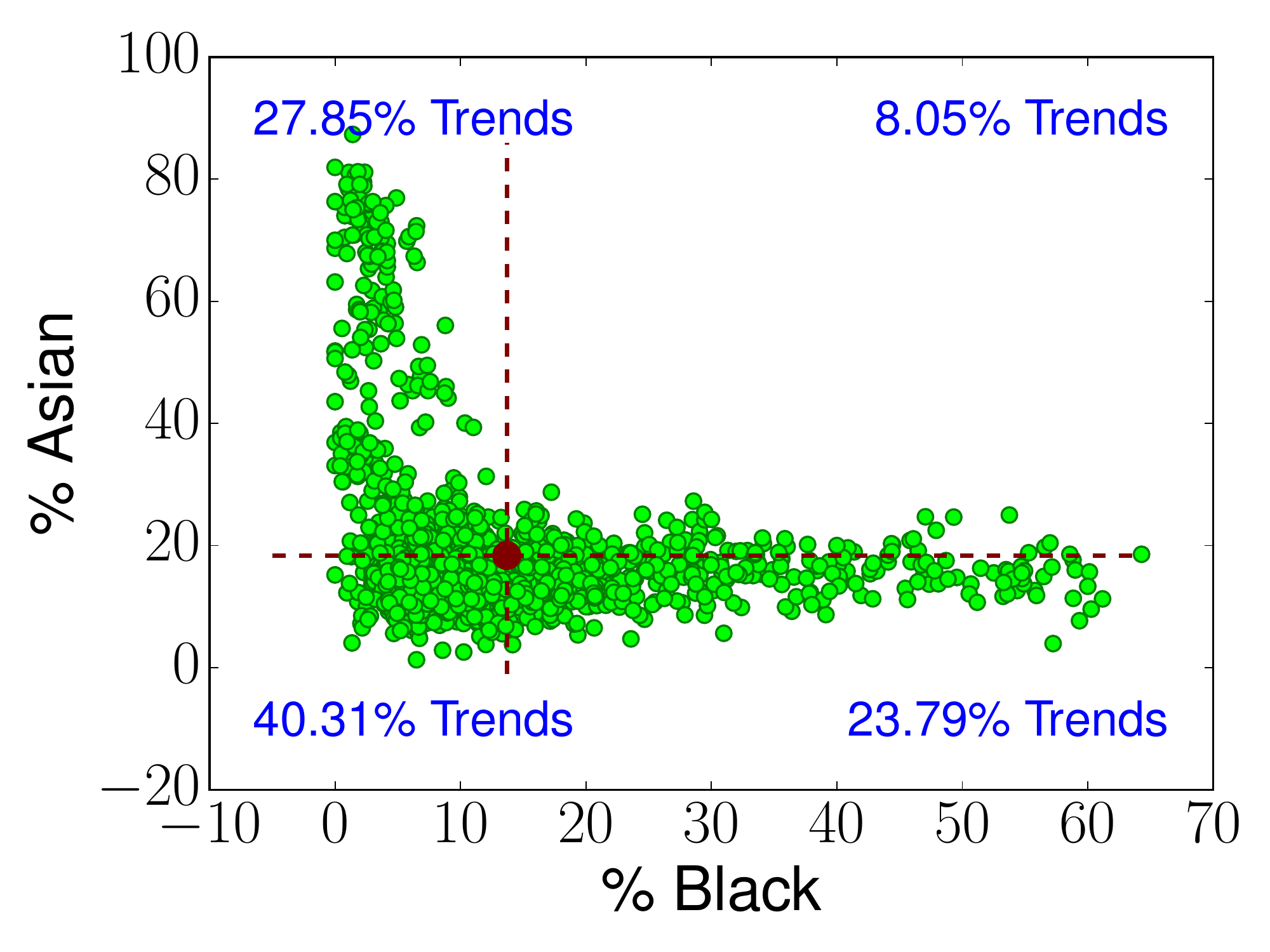}}
}
\vspace*{-2mm}
\caption{{\bf Racial distribution among the promoters of Twitter trends. Green dots represent the proportion of (a) Whites and Blacks, (b) Whites and Asians, (c) Blacks and Asians among the promoters. The proportion of the other race in each of the three plots can be implicitly derived. Red dots represent the proportion of corresponding races (e.g., Whites and Blacks in (a)) in overall Twitter population. In each plot, two red scattered lines divide the trends into four quadrants: first quadrant containing the trends where both races are represented more among the promoters, third quadrant containing trends where both races are represented less, second and fourth quadrants containing trends where one of the races is represented more and the other is represented less compared to their share in overall Twitter population.}}
\label{fig:scatter_race}
\vspace*{-3mm}
\end{figure*}

\begin{figure*}[tb]
\center{
\subfloat[{\bf }]{\includegraphics[width=0.33\textwidth]{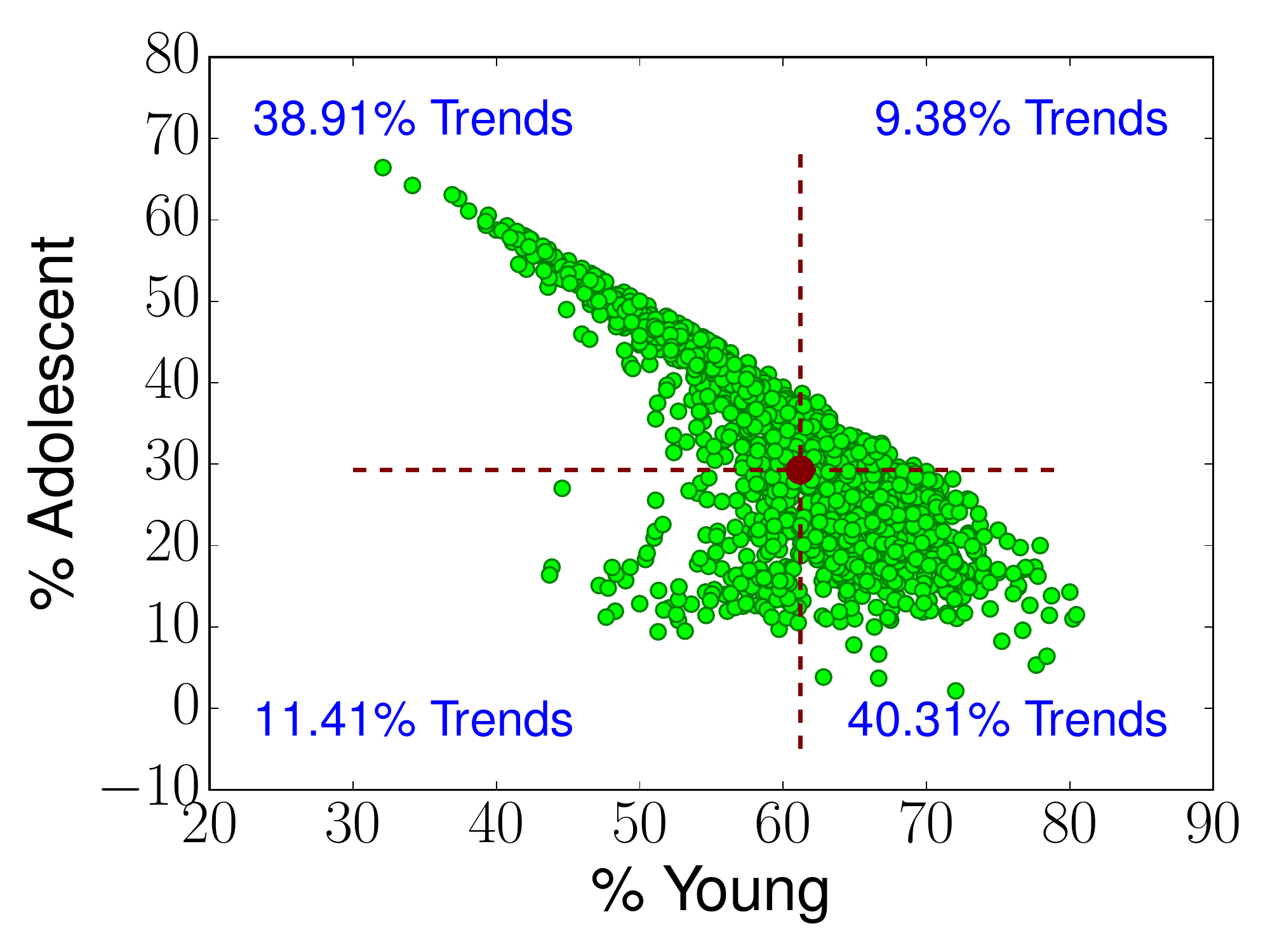}}
\hfil
\subfloat[{\bf }]{\includegraphics[width=0.33\textwidth]{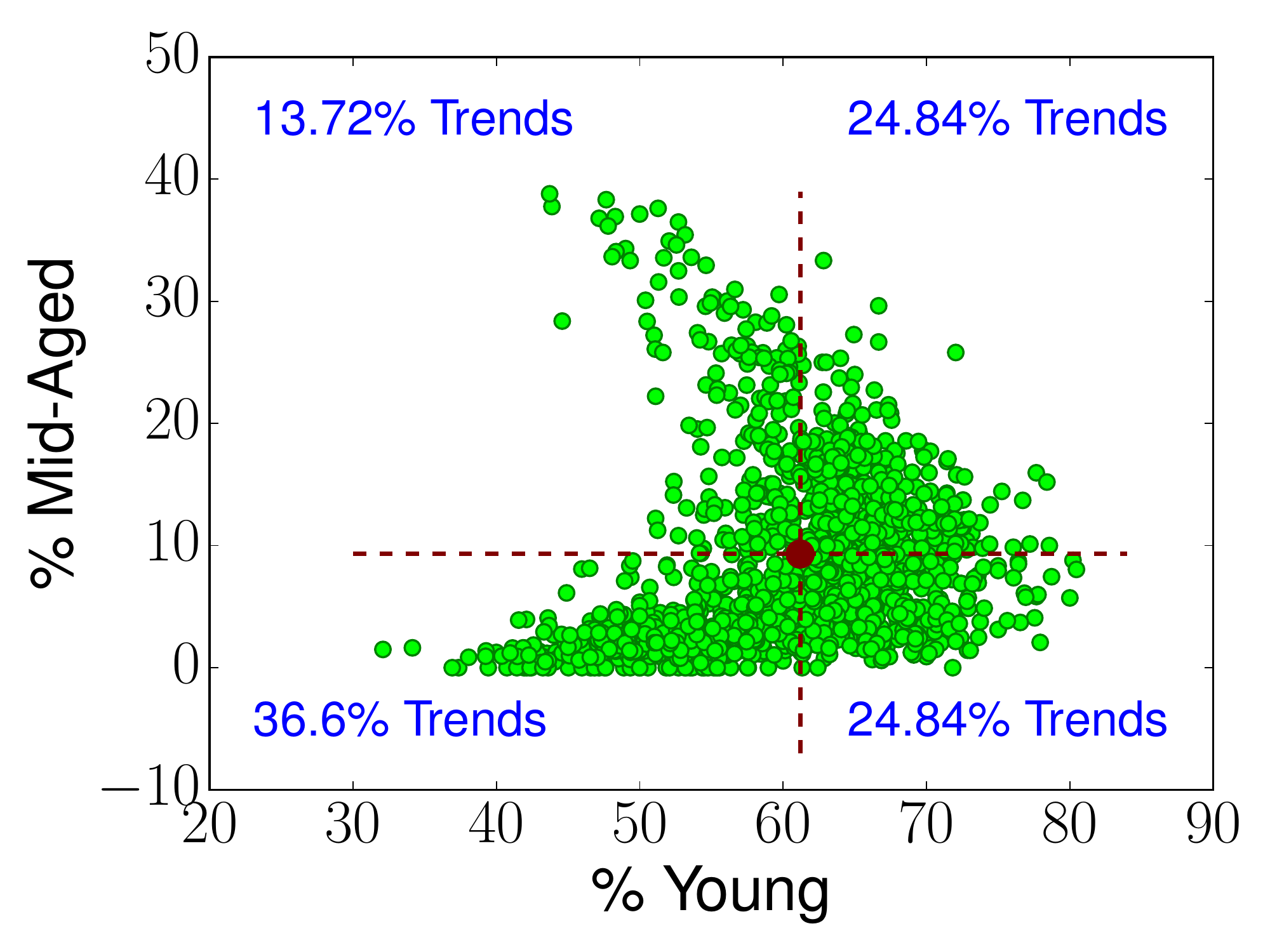}}
\hfil
\subfloat[{\bf }]{\includegraphics[width=0.33\textwidth]{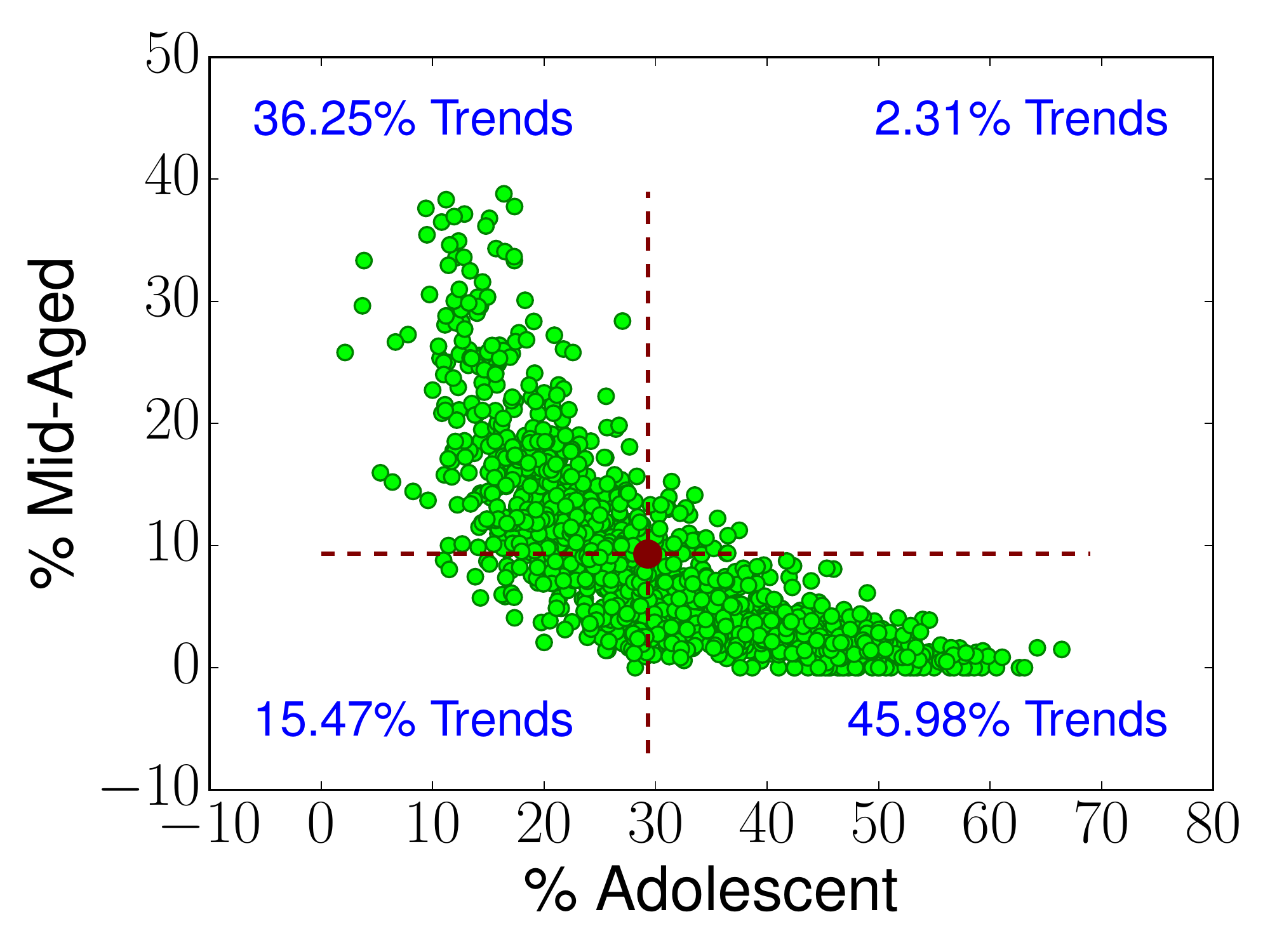}}
}
\vspace*{-2mm}
\caption{{\bf Distribution of age-groups among the promoters of Twitter trends. Green dots represent the proportion of (a) Young and Adolescents, (b) Young and Mid-aged people, (c) Mid-aged and Adolescents among the promoters. Red dot represent the proportion of corresponding age-groups in overall Twitter population. Similar to the Figures showing racial distributions, two red scattered lines divide the trends into four quadrants, where each quadrant contain trends where certain age-groups are represented more or represented less among the promoters.}}
\label{fig:scatter_age}
\vspace*{-2mm}
\end{figure*}

\begin{figure*}[tb]
\center{
\subfloat[{\bf }]{\includegraphics[width=0.33\textwidth]{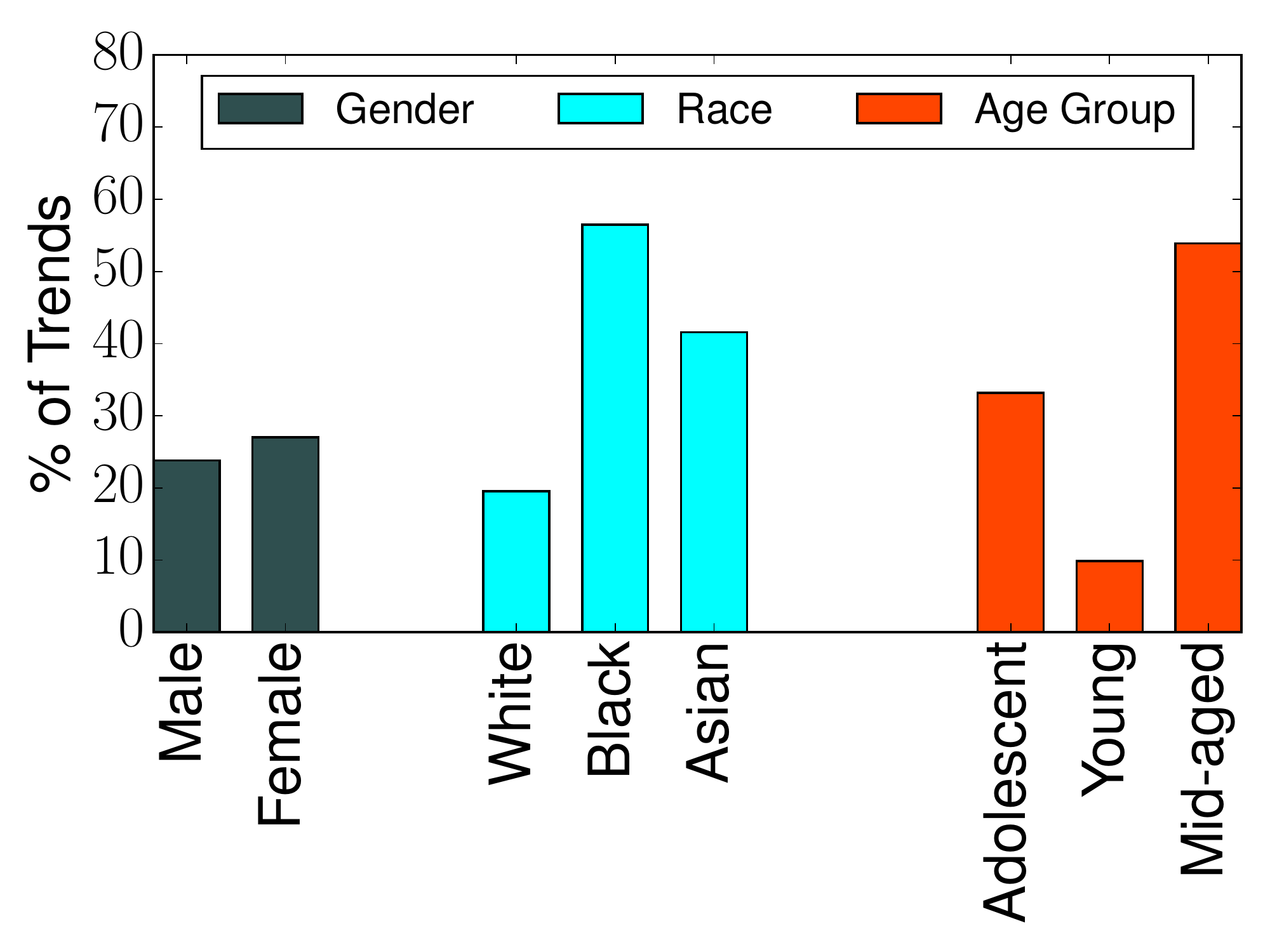}}
\hfil
\subfloat[{\bf }]{\includegraphics[width=0.33\textwidth]{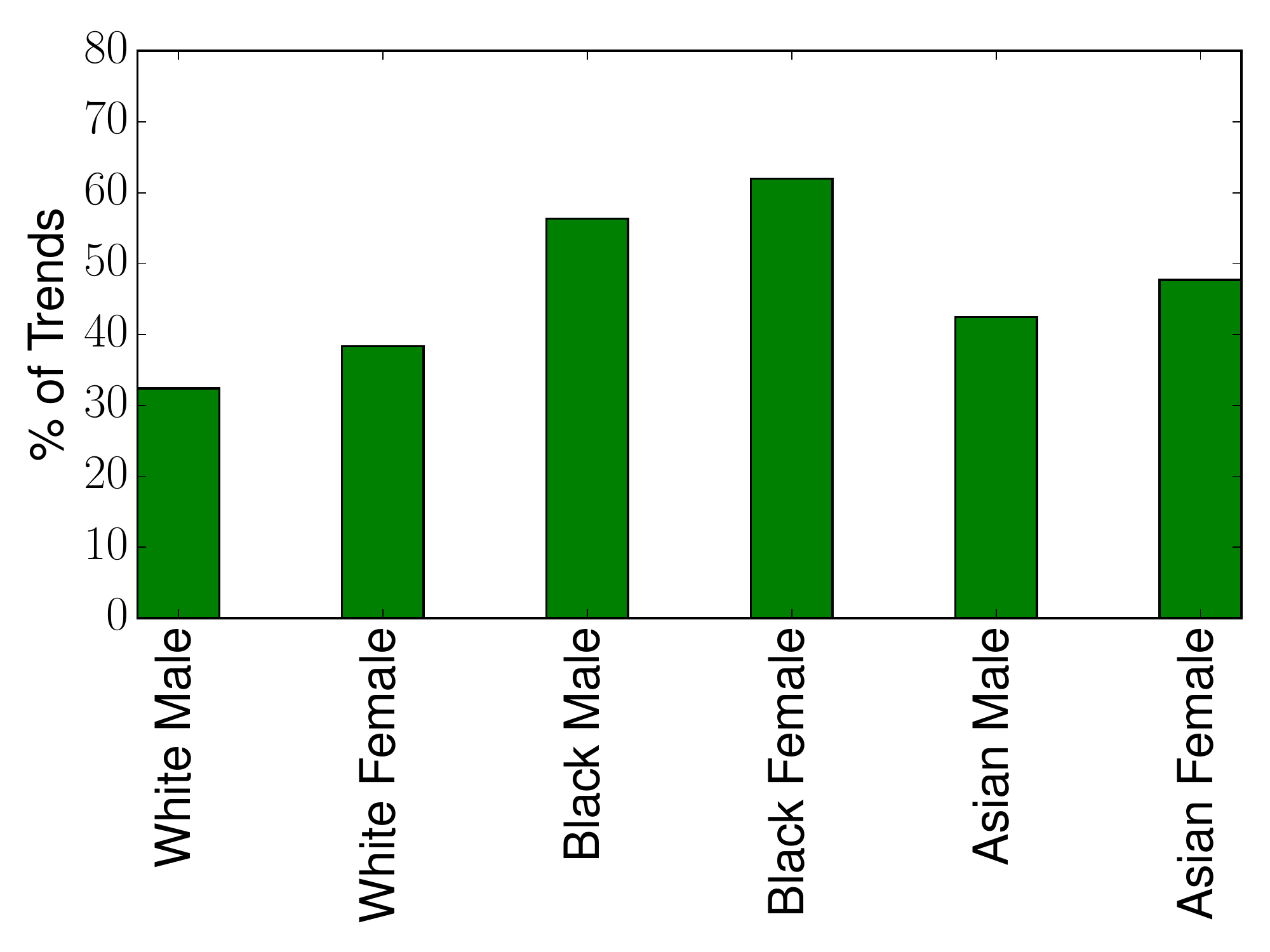}}
\hfil
\subfloat[{\bf }]{\includegraphics[width=0.33\textwidth]{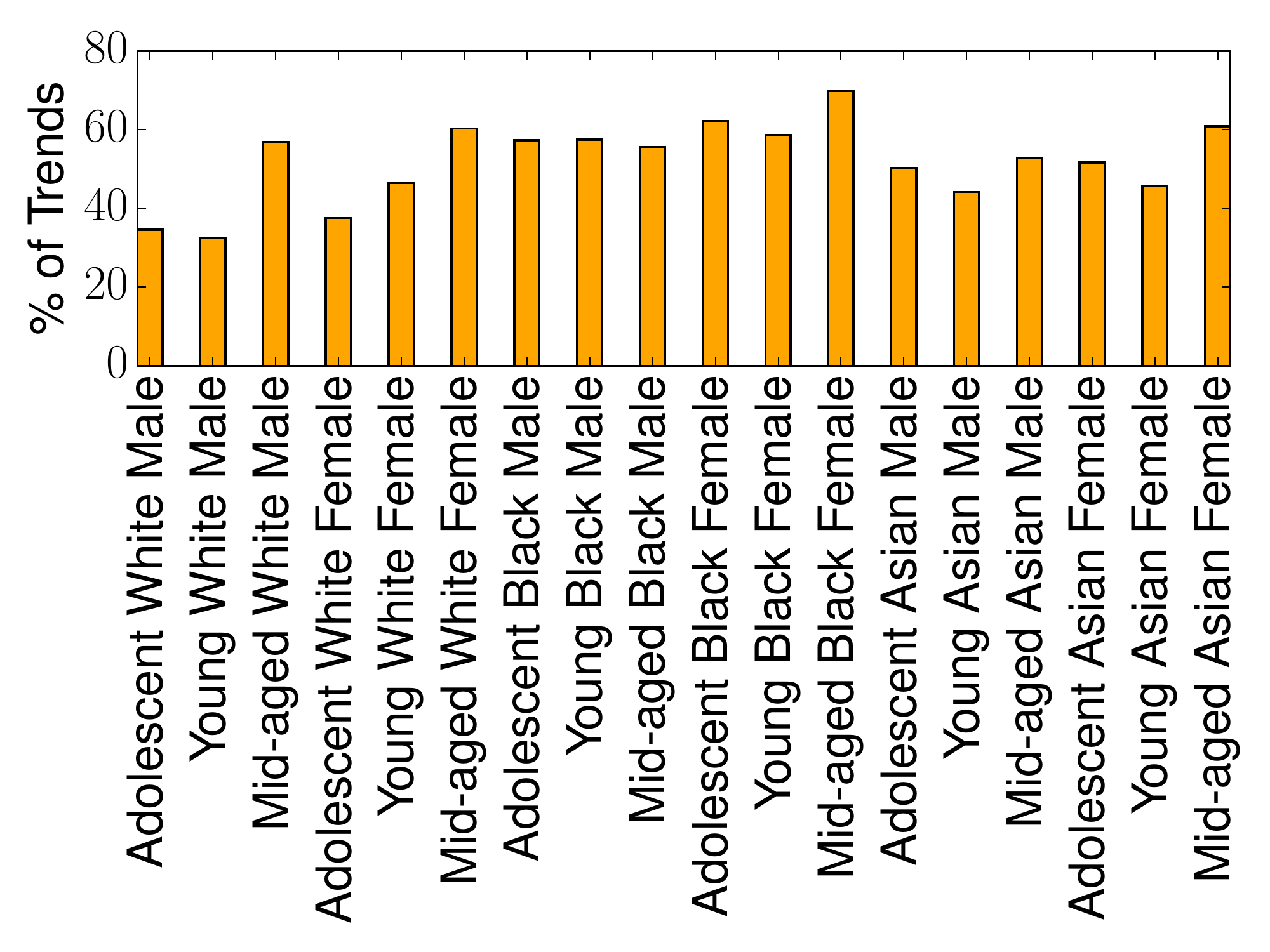}}
}
\vspace*{-2mm}
\caption{{\bf Percentage of trends where different demographic groups are {\it under-represented}: (a) considering gender, race and age independently, (b) considering both race and gender, and (c) considering all attributes together.}}
\label{fig:bias_ratio}
\vspace*{-2mm}
\end{figure*}

\subsection{Demographics of the promoters of Twitter trends}
\noindent
After computing the demographics of the Twitter population in US,
we now investigate the demographics
of the promoters of different trending topics.
For our analysis, we only consider $1,430$ trends from our dataset,
where we have the demographic information of at least $100$ of
their promoters.
We compute the distribution of genders, races and age-groups
among the promoters of these trends.
Figure~\ref{fig:scatter_gender}, Figure~\ref{fig:scatter_race} and Figure~\ref{fig:scatter_age}
respectively show the scatter plots of gender distribution, racial distribution, and distribution of
age-groups among the trend promoters.

The figures reveal that the trending topics in Twitter are promoted by
users of varied demographics. For example, we can see in Figure~\ref{fig:scatter_gender}
that there are a number of trends where men dominate the group of promoters. However,
there are also trends which are promoted mostly by women.

We can make similar observations from the racial distribution of
promoters in Figures~\ref{fig:scatter_race}(a), ~\ref{fig:scatter_race}(b), and
~\ref{fig:scatter_race}(c), as well as from the distribution of age-groups
in Figures~\ref{fig:scatter_age}(a), ~\ref{fig:scatter_age}(b), and
~\ref{fig:scatter_age}(c). In these figures, the scatter plots tend to form triangles,
where the boundaries (or edges) of the triangles represent extreme trends where
users from a particular demographic group (e.g., Blacks or Mid-aged people)
are not at all present among the promoters.

\begin{table*}[t]
\center
\small
\begin{tabular}{|p{0.18\textwidth}|p{0.06\textwidth}|p{0.045\textwidth}|p{0.045\textwidth}|p{0.12\textwidth}|p{0.115\textwidth}|p{0.09\textwidth}|p{0.16\textwidth}|}
\hline
{\bf Demographic Attribute} & {\bf Gender} & {\bf Race} & {\bf Age} & {\bf Gender \& Race} & {\bf Gender \& Age} & {\bf Race \& Age} & {\bf Gender, Race \& Age} \\
\hline
{\bf \% of Trends}  & 61.23 & 80.19 & 76.54 & 82.44 & 83.07 & 83.91 & 88.25 \\
\hline
\end{tabular}
\vspace*{-2mm}
\caption{{\bf Percentage of trends where demographics of the promoters differ {\it significantly} from overall Twitter population.}}
\label{tab:significantly_different}
\vspace*{-5mm}
\end{table*}

\subsection{Divergence of trend promoters from overall population}
\noindent
While analyzing the demographics of the promoters of different trends, we observed that
different trends are promoted by user-groups having highly disparate demographics.
Now, we compare their demographics with the (baseline or reference)
demographics of overall Twitter population in the US.
In Figure~\ref{fig:scatter_gender}, Figure~\ref{fig:scatter_race} and Figure~\ref{fig:scatter_age},
we present the baseline as a red circle. We observe that although some of the green circles,
representing demographics of the promoters, are close to the red circle, many of them are far from it.

To formally quantify whether the demographics of the promoters of a trend
deviates {\it significantly} from the demographics of the overall population,
we use {\bf Fisher's Exact Test}~\cite{fisher1922interpretation}.
For example, considering the gender demographics, if the
number of men and women among the promoters of a trend $t$
and in the overall population are
$N_1$, $N_2$, $N_3$ and $N_4$ respectively, then to evaluate
whether the proportion of men and women among the promoters of $t$
is significantly different from their proportion in overall population,
we first build the following {\it Contingency Table}
$$
\left| \begin{array}{c c}
N_1 & N_2 \\
N_3 & N_4
\end{array} \right|
$$
Then, we compute the {\it p-value} from this contingency table using Fisher's test~\cite{fisher1922interpretation}.
If the {\it p-value} obtained from the test is less than $0.05$,
we conclude that the difference in the two proportions
is {\it statistically significant}.
Although Fisher's exact test was originally proposed for $2 \times 2$
contingency tables, it has later been extended to apply on general
$r \times c$ contingency tables~\cite{mehta1983network}.

Table~\ref{tab:significantly_different} shows the fraction of trends,
which are promoted by groups of users
who are {\it significantly} different from Twitter's overall user population.
We can see that such trends constitute a significant majority of all trending topics,
which indicates that the promoters of most of the trends are different from the
overall population.

This observation is interesting because when a topic is declared trending on Twitter,
and news outlets start reporting on
them\footnote{\it fortune.com/2017/02/21/delete-uber-twitter}
\footnote{\it teenvogue.com/story/day-without-immigrants-strike-twitter},
the underlying assumption is that the topic is popular among a {\it representative sample
of all Twitter users} in a geographical region.
However, as we see in our analysis, this assumption does not hold in practice.
Hence, along with the topic, it is also important to know the specific groups of users
who make the topic trending.

When the representation of different demographic groups
(such as Whites, Women, or Adolescents) among the trend promoters
deviate from the overall Twitter population,
the groups can either be {\it represented more} or {\it represented less}
compared to their share in the overall population.
To investigate how different groups are represented,
we plot reference lines along the x-axis in Figure~\ref{fig:scatter_gender},
and along both x-axis and y-axis in
Figures~\ref{fig:scatter_race}(a-c) and Figures~\ref{fig:scatter_age}(a-c).
Each reference line denotes the proportion of users in the overall Twitter
population belonging to a particular demographic group.
For example, the reference line in Figure~\ref{fig:scatter_gender} denotes
the percentage of men among the overall Twitter population.
This line divides the trends in Figure~\ref{fig:scatter_gender}
into two halves: (i) upper-half, which contains the trends where men are represented
more among the promoters
(there are $53.25\%$ of all trends falling into the upper half),
and (ii) lower-half containing trends ($46.75\%$) where men are represented less.

For Figures~\ref{fig:scatter_race}(a-c) and Figures~\ref{fig:scatter_age}(a-c),
the reference lines divide the trends
into four quadrants: (i) first quadrant contains the trends where both demographic groups shown in a particular figure
are represented more among the promoters; (iii) third quadrant contains trends where both groups
are represented less; (ii) second quadrant and (iv) fourth quadrant have the trends where one of the
groups is represented more and the other is represented less among the promoters,
compared to the overall Twitter population.

\subsection{Under-representation of demographic groups}
\noindent
For most of the trends, we observed that different demographic groups
(such as Whites, Women, or Adolescents)
are represented less or represented more among the promoters of these trends. 
%, compared to their representations in the overall Twitter population.
A pertinent question to ask in this context is for how many trends,
the representation of a particular demographic group is {\it significantly less}.

We define a demographic group $j$ to be {\it under-represented} among the promoters of
 a topic, if the fraction of promoters belonging to group $j$ is less than $80\%$ of the fraction of $j$ in the
reference demographics (i.e., the overall population). Our selection of the $80\%$ threshold is motivated
by the {\it 80\% rule} used by U.S. Equal Employment Opportunity Commission to determine
whether a company's hiring policy has any {\it adverse impact} on a minority group~\cite{biddle2006adverse}.

Figure~\ref{fig:bias_ratio}(a) shows the under-representation of different gender groups, racial groups and age groups.
Figure~\ref{fig:bias_ratio}(b) and Figure~\ref{fig:bias_ratio}(c) respectively show the under-representation for the demographic groups
based on both race and gender, as well as based on all three demographic attributes together.
The age group `old' is not shown in these figures as we do not have enough tweets posted by old people
in our dataset.

We make the following interesting observations in Figures~\ref{fig:bias_ratio}(a),~\ref{fig:bias_ratio}(b),
and~\ref{fig:bias_ratio}(c): \\
(i) Although the fraction of women in the Twitter population is larger than that of men, women are
under-represented more among the trend promoters than men. \\
(ii) Blacks and Asians are under-represented in the racial demographic groups promoting Twitter trends.\\
(iii) Among the age-groups, both adolescents and mid-aged people are under-represented.\\
(iv) Considering race and gender together,
among all groups, black women are most under-represented. \\
(v) Among the demographic groups based on gender, race and age together,
the highest under-representation is noticed in mid-aged black women.

Our observations about the perceived under-representation towards women and
the demographic groups containing black users are in line with previous findings related to
gender inequalities in Twitter~\cite{nilizadeh2016twitter}, and in many other efforts that discuss
inequalities of these demographic groups in our society~\cite{cotter2001glass,bonilla2006racism}.
More importantly, these observations suggest that the so called
`glass ceiling effect', usually used to describe the barriers that women face at the highest levels of
an organization~\cite{cotter2001glass}, may occur even in crowdsourced recommendations such as Twitter Trends.

\begin{figure}[t]
\center{
\includegraphics[width=0.7\columnwidth]{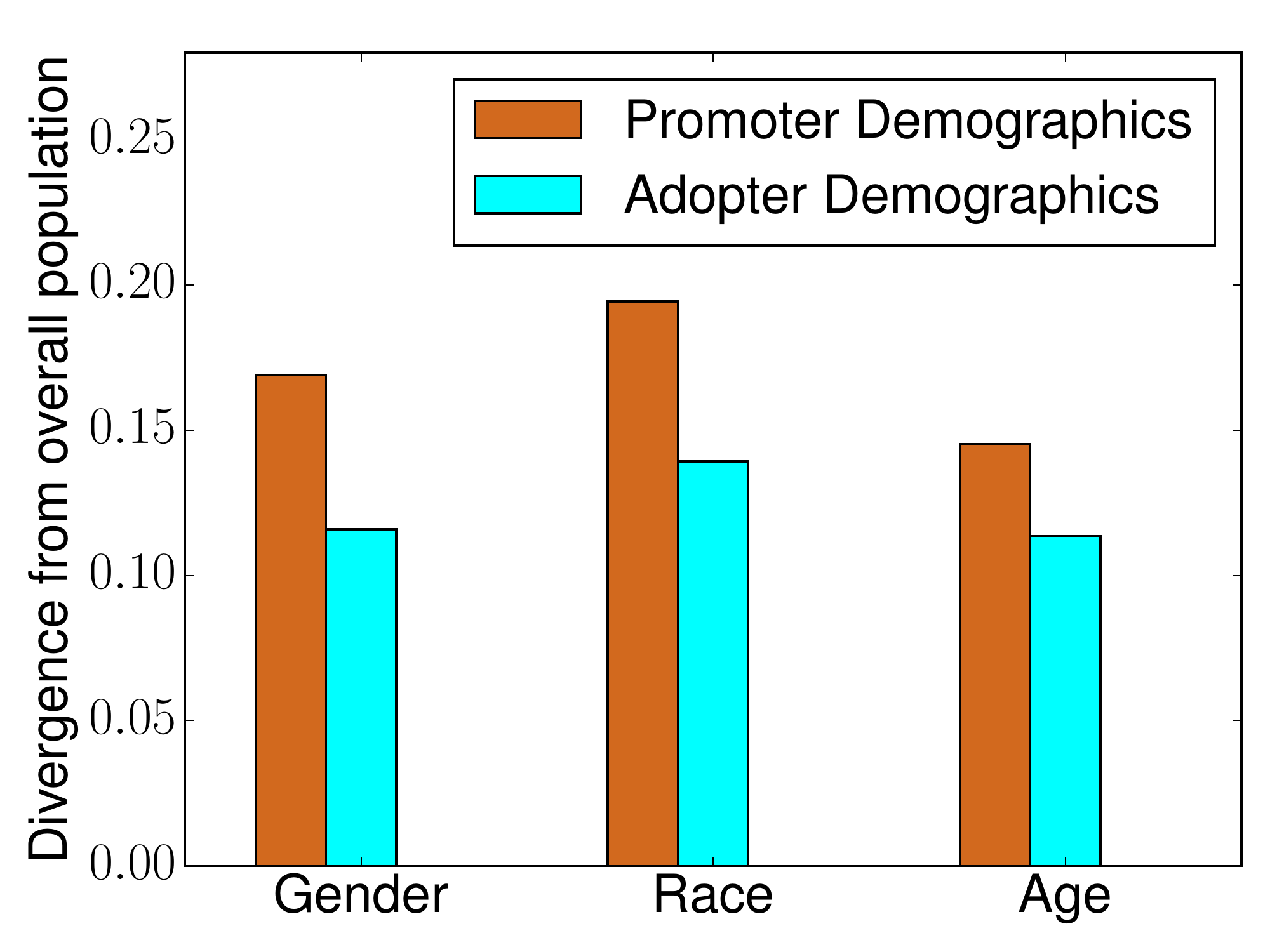}
}
\caption{{\bf Divergence of trend promoters and adopters from the overall Twitter population.}}
\label{fig:promoter_adopter_bias}
\vspace*{-5mm}
\end{figure}

\subsection{Impact of trend recommendations}
\noindent
We next turn our focus towards the {\it adopters} of the trends who have used the topics
{\it after} they became trending. It is expected that once a topic becomes trending,
it gets noticed by a large number of users.
We want to investigate whether the topics still remain popular among the same demographics who promoted them
(before they became trending),
or whether the topics get adopted by a wider population.
Towards that end, similar to the demographics of promoters, we compute the
demographics of adopters for all trends by considering the proportion of adopters belonging
to different demographic groups, and then consider their divergence from the overall Twitter demographics.

While computing the {\it divergence} of demographics of the users of a topic $i$ (either promoting or adopting it)
from the {\it reference} or {\it baseline} demographics of all Twitter users, we consider the euclidean distance between
the demographics of the promoters (and adopters) of $i$ and the reference demographics:
\begin{equation}
Divergence_{i} = || d_i - d_r ||
\end{equation}
where $d_r$ is the reference or baseline demographics.
Higher the score, more divergent is the demographics of users promoting or adopting the topic.

Figure~\ref{fig:promoter_adopter_bias} shows the average divergence of promoters and adopters of different trends from the overall population.
We can see in Figure~\ref{fig:promoter_adopter_bias} that the demographics of the adopters
 for different trends become closer to the overall Twitter population, and thereby reducing the divergence score. 
 Thus, we can conclude that the topics which were promoted by particular sections of the users, get adopted by a much wider population after the topics become trending.

\begin{table*}[t]
\center
\small
\begin{tabular}{|p{0.18\textwidth}|p{0.16\textwidth}|p{0.58\textwidth}|}
\hline
{\bf Demographic Attribute} & {\bf Demographic Group} & {\bf Example Trends} \\
\hline
\multirow{2}{*}{Gender} & {Male} & \#SXSW2017, \#Wikileaks, \#HeGone, \#playstationmeeting, \#drunkerhistory  \\
\cline{2-3}
~ & Female & \#JUSTINFOREVER, \#BacheloretteFinale, \#mypetmystar, \#weloveyounormani   \\
\hline
\multirow{3}{*}{Race} & {White} & \#PardonSnowden, \#UnlikelyBreakfastCereals, \#PrisonStrike, \#NightTube\\
\cline{2-3}
~ & Black & \#OneAfricaMusicFest, \#blackloveday, \#BlackGirlsRock, \#ThingsBlackpplFear  \\
\cline{2-3}
~ & Asian & \#FlyInNYC, \#ButterflyKiss, \#indiedevhour, \#QueenHwasaDay \\
\hline
\multirow{3}{*}{Age-group} & {Adolescent} & \#WhoSaysYouAreNotPerfectSelena, \#NationalTeddyBearDay, \#SuperJunior \\
\cline{2-3}
~ & Young & \#breastfeeding, \#KeepItPersonal, \#WWEChicago \\
\cline{2-3}
~ & Mid-aged & \#HillaryHealth, \#TrumpInDetroit, \#WheresYourTaxes, \#TrumpCouldSay \\
\hline
\end{tabular}
\vspace*{-2mm}
\caption{{\bf Examples of trends that are promoted by mostly one demographic group.}}
\label{tab:niche_interest}
\vspace*{-3mm}
\end{table*}

\begin{table*}[t]
\center
\small
\begin{tabular}{|p{0.2\textwidth}|p{0.056\textwidth}|p{0.075\textwidth}||p{0.062\textwidth}|p{0.062\textwidth}|p{0.062\textwidth}||p{0.1\textwidth}|p{0.07\textwidth}|p{0.09\textwidth}|}
\hline
\multirow{3}{*}{\bf Trend} & \multicolumn{8}{c|}{\bf Demographics of Promoters} \\
\cline{2-9}
~ & \multicolumn{2}{c||}{\bf Gender}  & \multicolumn{3}{c||}{\bf Race} & \multicolumn{3}{c|}{\bf Age-group} \\
\cline{2-9}
~ & \% Male & \% Female & \% White & \% Black & \% Asian & \% Adolescent & \% Young & \% Mid-aged \\
\hline
\#AmericaWasNeverGreat & 47.5 & 52.5 &  \textcolor{red}{\it 63.13} & \textcolor{blue}{\bf 21.25} & \textcolor{red}{\it 15.63} &  \textcolor{red}{\it 28.75} & \textcolor{red}{\it 59.38} & \textcolor{blue}{\bf 11.25}  \\
%\hline
%\#FourthOfJuly & \textcolor{blue}{\bf 48.79} & \textcolor{red}{\it 51.21} &  \textcolor{blue}{\bf 76.6} & \textcolor{red}{\it 10.13} & \textcolor{red}{\it 13.27} &  \textcolor{red}{\it 22.54} & \textcolor{blue}{\bf 64.91} & \textcolor{blue}{\bf 12.13}  \\
\hline
\#Freedom & \textcolor{blue}{\bf 54.01} & \textcolor{red}{\it 45.99} &  \textcolor{blue}{\bf 80.21} & \textcolor{red}{\it 8.02} & \textcolor{red}{\it 11.76} &  \textcolor{red}{\it 20.86} & \textcolor{blue}{\bf 65.78} & \textcolor{blue}{\bf 12.3}  \\
\hline
\#GodBlessAmerica & 48.97 & 51.03 &  \textcolor{blue}{\bf 80.69} & \textcolor{red}{\it 10.34} & \textcolor{red}{\it 8.97} &  \textcolor{red}{\it 17.24} & \textcolor{blue}{\bf 64.83} & \textcolor{blue}{\bf 17.93}  \\
\hline
\#GrowingUpInABlackChurch & \textcolor{red}{\it 37.02} & \textcolor{blue}{\bf 62.98} &  \textcolor{red}{\it 32.75} & \textcolor{blue}{\bf 53.29} & \textcolor{red}{\it 13.95} &  \textcolor{blue}{\bf 45.93} & \textcolor{red}{\it 50.58} & \textcolor{red}{\it 3.49}  \\
%\hline
%\#StopTheHate & \textcolor{red}{\it 45.65} & \textcolor{blue}{\bf 54.35} &  \textcolor{blue}{\bf 71.74} & \textcolor{red}{\it 6.52} & \textcolor{blue}{\bf 21.74} &  \textcolor{blue}{\bf 39.13} & \textcolor{red}{\it 52.17} & \textcolor{red}{\it 8.7}  \\
\hline
\#WeAreAmerica & \textcolor{blue}{\bf 62.07} & \textcolor{red}{\it 37.93} &  \textcolor{red}{\it 63.79} & \textcolor{blue}{\bf 18.97} & \textcolor{red}{\it 17.24} &  \textcolor{red}{\it 22.41} & \textcolor{blue}{\bf 67.24} & \textcolor{red}{\it 8.62}  \\
\hline
\end{tabular}
\vspace*{-2mm}
\caption{{\bf Demographics of promoters of Twitter trends during US Independence Day (4th July, 2016). Demographic groups shown in bold blue are represented more, and groups in red italics are represented less among the promoters.}}
\label{tab:independence_day}
\vspace*{-4mm}
\end{table*}

\begin{table*}[t]
\center
\small
\begin{tabular}{|p{0.2\textwidth}|p{0.056\textwidth}|p{0.075\textwidth}||p{0.062\textwidth}|p{0.062\textwidth}|p{0.062\textwidth}||p{0.1\textwidth}|p{0.07\textwidth}|p{0.09\textwidth}|}
\hline
\multirow{3}{*}{\bf Trend} & \multicolumn{8}{c|}{\bf Demographics of Promoters} \\
\cline{2-9}
~ & \multicolumn{2}{c||}{\bf Gender}  & \multicolumn{3}{c||}{\bf Race} & \multicolumn{3}{c|}{\bf Age-group} \\
\cline{2-9}
~ & \% Male & \% Female & \% White & \% Black & \% Asian & \% Adolescent & \% Young & \% Mid-aged \\
\hline
\#AllLivesShouldMatter & 47.37 & 52.63 &  \textcolor{red}{\it 47.36} & \textcolor{blue}{\bf 26.32} & \textcolor{blue}{\bf 26.32} &  \textcolor{blue}{\bf 63.12} & \textcolor{red}{\it 36.8} & \textcolor{red}{\it 0.8} \\
\hline
\#BattleBots & \textcolor{blue}{\bf 66.67} & \textcolor{red}{\it 33.33} &  69.44 & 11.11 & 19.44 &  \textcolor{red}{\it 5.56} & \textcolor{blue}{\bf 72.22} & \textcolor{blue}{\bf 22.22} \\
\hline
\#BlackLivesMatter & 50.0 & 50.0 &  \textcolor{red}{\it 28.57} & \textcolor{blue}{\bf 57.14} & \textcolor{red}{\it 14.29} &  \textcolor{red}{\it 21.43} & \textcolor{blue}{\bf 64.29} & \textcolor{blue}{\bf 14.29}  \\
\hline
\#BlackSkinIsNotACrime & \textcolor{red}{\it 43.86} & \textcolor{blue}{\bf 56.14} &  \textcolor{red}{\it 38.6} & \textcolor{blue}{\bf 38.6} & \textcolor{blue}{\bf 22.81} &  \textcolor{blue}{\bf 43.86} & \textcolor{red}{\it 54.39} & \textcolor{red}{\it 1.75}  \\
\hline
\#DallasPoliceShootings & 45.18 & 54.82 &  \textcolor{blue}{\bf 77.71} & \textcolor{red}{\it 9.64} & \textcolor{red}{\it 12.65} &  \textcolor{red}{\it 28.31} & \textcolor{red}{\it 54.82} & \textcolor{blue}{\bf 15.66}  \\
\hline
%\#FalconHeightsShooting & \textcolor{red}{\it 38.4} & \textcolor{blue}{\bf 61.6} &  \textcolor{red}{\it 49.74} & \textcolor{blue}{\bf 29.01} & \textcolor{blue}{\bf 21.25} &  \textcolor{blue}{\bf 40.53} & \textcolor{red}{\it 54.1} & \textcolor{red}{\it 5.29}  \\
%\hline
\#PoliceLivesMatter & 50.0 & 50.0 &  \textcolor{blue}{\bf 73.68} & 13.16 & \textcolor{red}{\it 13.16} &  \textcolor{red}{\it 15.79} & \textcolor{blue}{\bf 65.79} & \textcolor{blue}{\bf 18.42}  \\
%\hline
%\#PrayForAmerica & \textcolor{red}{\it 42.11} & \textcolor{blue}{\bf 57.89} &  \textcolor{red}{\it 61.99} & \textcolor{blue}{\bf 16.96} & \textcolor{blue}{\bf 21.05} &  \textcolor{blue}{\bf 38.6} & \textcolor{red}{\it 54.39} & \textcolor{red}{\it 7.02}  \\
\hline
\#PrayForPeace & \textcolor{red}{\it 35.14} & \textcolor{blue}{\bf 64.86} &  \textcolor{blue}{\bf 78.38} & \textcolor{red}{\it 2.7} & 18.92 &  29.73 & \textcolor{blue}{\bf 64.86} & \textcolor{red}{\it 5.41}  \\
%\hline
%\#StopTheViolence & \textcolor{blue}{\bf 47.08} & \textcolor{red}{\it 52.92} &  \textcolor{red}{\it 67.01} & \textcolor{blue}{\bf 17.53} & \textcolor{red}{\it 15.46} & \textcolor{blue}{\bf 32.65} & \textcolor{red}{\it 60.14} & \textcolor{red}{\it 7.22} \\
\hline
\end{tabular}
\vspace*{-2mm}
\caption{{\bf Demographics of promoters of different Twitter trends during Dallas Shooting (7th and 8th July, 2016). Demographic groups shown in bold blue are represented more, and groups shown in red italics are represented less among the promoters.}}
\label{tab:dallas_shooting}
\vspace*{-3mm}
\end{table*}

\begin{table*}[t]
\center
\small
\begin{tabular}{|p{0.2\textwidth}|p{0.056\textwidth}|p{0.075\textwidth}||p{0.062\textwidth}|p{0.062\textwidth}|p{0.062\textwidth}||p{0.1\textwidth}|p{0.07\textwidth}|p{0.09\textwidth}|}
\hline
\multirow{3}{*}{\bf Trend} & \multicolumn{8}{c|}{\bf Demographics of Promoters} \\
\cline{2-9}
~ & \multicolumn{2}{c||}{\bf Gender}  & \multicolumn{3}{c||}{\bf Race} & \multicolumn{3}{c|}{\bf Age-group} \\
\cline{2-9}
~ & \% Male & \% Female & \% White & \% Black & \% Asian & \% Adolescent & \% Young & \% Mid-aged \\
\hline
\hline
\multicolumn{9}{|c|}{\bf 7th November, 2016} \\
%\hline
%\#ImVotingBecause & \textcolor{blue}{\bf 53.36} & \textcolor{red}{\it 46.64} &  \textcolor{blue}{\bf 81.67} & \textcolor{red}{\it 8.35} & \textcolor{red}{\it 9.98} &  \textcolor{red}{\it 19.95} & \textcolor{red}{\it 58.24} & \textcolor{blue}{\bf 20.65}  \\
\hline
\#ImWithHer & 44.44 & 55.56  & \textcolor{blue}{\bf 77.78} & 11.11 & \textcolor{red}{\it 11.11} &  \textcolor{red}{\it 22.22} & \textcolor{blue}{\bf 66.67} & \textcolor{blue}{\bf 11.11} \\
\hline
\#MyVote2016 & \textcolor{blue}{\bf 71.43} & \textcolor{red}{\it 28.57} &  \textcolor{blue}{\bf 82.21} & \textcolor{blue}{\bf 14.29} & \textcolor{red}{\it 3.5} &  \textcolor{red}{\it 14.29} & \textcolor{blue}{\bf 82.33} & \textcolor{red}{\it 3.38} \\
\hline
\#TrumpWinsBecause & \textcolor{blue}{\bf 78.57} & \textcolor{red}{\it 21.43} &  \textcolor{blue}{\bf 91.79} & \textcolor{red}{\it 1.07} & \textcolor{red}{\it 7.14} &  \textcolor{red}{\it 14.29} & \textcolor{red}{\it 42.86} & \textcolor{blue}{\bf 35.71}  \\
%\hline
%\#VoteProLife & \textcolor{red}{\it 44.44} & \textcolor{blue}{\bf 55.56} &  \textcolor{blue}{\bf 86.79} & \textcolor{red}{\it 11.11} & \textcolor{red}{\it 2.1} &  \textcolor{red}{\it 22.22} & \textcolor{red}{\it 55.56} & \textcolor{blue}{\bf 22.22} \\
\hline
\hline
\multicolumn{9}{|c|}{\bf 8th November, 2016} \\
\hline
\#Decision2016 & 44.24 & 55.76 &  \textcolor{blue}{\bf 82.49} & \textcolor{red}{\it 6.45} & \textcolor{red}{\it 11.06} &  \textcolor{red}{\it 22.58} & \textcolor{blue}{\bf 68.2} & \textcolor{red}{\it 9.22}  \\
\hline
\#ElectionDay & 50.46 & 49.54 &  \textcolor{blue}{\bf 77.52} & \textcolor{red}{\it 10.44} & \textcolor{red}{\it 12.04} &  \textcolor{red}{\it 20.88} & \textcolor{blue}{\bf 63.37} & \textcolor{blue}{\bf 15.16} \\
\hline
\#ObamaDay & 51.14 & 48.86 &  \textcolor{red}{\it 50.0} & \textcolor{blue}{\bf 36.36} & \textcolor{red}{\it 13.64} &  \textcolor{red}{\it 19.32} & \textcolor{blue}{\bf 71.59} & \textcolor{red}{\it 9.09} \\
\hline
\hline
\multicolumn{9}{|c|}{\bf 9th November, 2016} \\
\hline
\#ElectionResults & 49.57 & 50.43 &  \textcolor{blue}{\bf 81.2} & \textcolor{red}{\it 7.69} & \textcolor{red}{\it 11.11} &  \textcolor{red}{\it 13.68} & \textcolor{blue}{\bf 69.23} & \textcolor{blue}{\bf 17.09} \\
\hline
\#ImStillWithHer & \textcolor{red}{\it 31.17} & \textcolor{blue}{\bf 68.83} &  \textcolor{blue}{\bf 77.27} & \textcolor{red}{\it 9.09} & \textcolor{red}{\it 13.64} &  \textcolor{blue}{\bf 33.77} & \textcolor{red}{\it 57.79} & \textcolor{red}{\it 8.44}  \\
\hline
\#MorningAfter & \textcolor{red}{\it 38.46} & \textcolor{blue}{\bf 61.54} & 69.23 & \textcolor{red}{\it 7.69} & \textcolor{blue}{\bf 23.08} &  \textcolor{red}{\it 7.69} & \textcolor{blue}{\bf 76.92} & \textcolor{blue}{\bf 15.38}  \\
%\hline
%\#NotMyPresident & 46.68 & 53.32 &  \textcolor{blue}{\bf 72.31} & \textcolor{blue}{\bf 13.92} & \textcolor{red}{\it 13.77} &  \textcolor{blue}{\bf 31.33} & \textcolor{red}{\it 54.75} & \textcolor{blue}{\bf 12.97}  \\
\hline
\#NowThatTrumpIsPresident & \textcolor{blue}{\bf 76.47} & \textcolor{red}{\it 23.53} &  \textcolor{red}{\it 35.29} & \textcolor{blue}{\bf 41.18} & \textcolor{blue}{\bf 23.53} &  \textcolor{red}{\it 17.65} & \textcolor{blue}{\bf 70.59} & \textcolor{blue}{\bf 11.76} \\
%\hline
%\#OutNumbered & \textcolor{blue}{\bf 57.14} & \textcolor{red}{\it 42.86} &  \textcolor{blue}{\bf 71.43} & \textcolor{blue}{\bf 14.29} & \textcolor{red}{\it 14.29} &  \textcolor{red}{\it 28.57} & \textcolor{red}{\it 42.86} & \textcolor{blue}{\bf 28.57} \\
\hline
\#PresidentTrump & 47.47 & 52.53 &  \textcolor{blue}{\bf 77.22} & \textcolor{red}{\it 8.23} & \textcolor{red}{\it 14.56} &  \textcolor{red}{\it 18.35} & \textcolor{blue}{\bf 65.82} & \textcolor{blue}{\bf 14.56}  \\
%\hline
%\#ProudToBeCanadian & \textcolor{red}{\it 18.18} & \textcolor{blue}{\bf 81.82} &  \textcolor{blue}{\bf 72.73} & \textcolor{red}{\it 9.09} & \textcolor{red}{\it 18.18} &  \textcolor{red}{\it 27.27} & \textcolor{blue}{\bf 72.73} & \textcolor{red}{\it 0.0} \\
\hline
\#RIPAmerica & \textcolor{red}{\it 39.06} & \textcolor{blue}{\bf 60.94} &  \textcolor{blue}{\bf 73.44} & \textcolor{red}{\it 6.25} & \textcolor{blue}{\bf 20.31} &  \textcolor{blue}{\bf 40.63} & \textcolor{red}{\it 54.69} & \textcolor{red}{\it 4.69} \\
\hline
\#TrumpsFirstOrder & \textcolor{blue}{\bf 78.13} & \textcolor{red}{\it 21.88} &  \textcolor{blue}{\bf 84.38} & \textcolor{red}{\it 6.25} & \textcolor{red}{\it 9.38} &  \textcolor{red}{\it 15.63} & \textcolor{red}{\it 56.25} & \textcolor{blue}{\bf 28.13}  \\
\hline
\end{tabular}
\vspace*{-2mm}
\caption{{\bf Demographics of promoters of different Twitter trends during US presidential election 2016. Trends during different days are listed separately. Demographic groups shown in bold blue are represented more, and groups shown in red italics are represented less among the promoters.}}
\label{tab:us_election}
\vspace*{-5mm}
\end{table*}

\subsection{Demographics influencing the type of trends}
Earlier, we saw how different demographic groups are represented  among the
promoters of different trends. We now attempt to analyze the impact of  the demographics of the promoters on the type of topics becoming trending. Looking through the trends promoted by users where a particular demographic group is represented more, we find two patterns among such trends: (i) the trends tend to express the niche interests of that demographic group, (ii) when some event happens, which is of interest to everyone, different trends bring out different perspectives on that event. Next, we demonstrate these observations with some case studies.

\vspace{1mm}
\noindent \textbf{Trends expressing niche interest} \\
We first look at some of the trends where the promoters are dominated by certain demographic groups, and list some examples in Table~\ref{tab:niche_interest}.

\vspace{1mm}
\noindent \textit{Trends promoted more by one gender group}\\
Second and third rows in Table~\ref{tab:niche_interest} show the trends promoted by either mostly men, or mostly women.
We see that the gaming trends like \#playstationmeeting, or political trends like \#Wikileaks, are mostly promoted by men.
On the other hand, trends \#JUSTINFOREVER, about the singer Justin Bieber and \#BacheloretteFinale, a TV show, are promoted mostly by women.

\vspace{1mm}
\noindent \textit{Trends promoted more by one racial group}\\
Fourth, fifth and sixth rows in Table~\ref{tab:niche_interest} show the trends promoted by either mostly Whites, Blacks, or  Asians.
Trends such as \#UnlikelyBreakfastCereals and \#PrisonStrike, which is related to the protest of prison inmates, are more popular among Whites.
Whereas, trends of niche cultural interest such as \#OneAfricaMusicFest, \#blackloveday are more popular among Blacks. Finally, trends popular mostly among Asians include \#FlyInNYC, which is related to
a concert by the South Korean musical group GOT7, and the trend \#QueenHwasaDay, which is about Hwasa,
a member of the Korean girl band namely Mamamoo.

\vspace{1mm}
\noindent \textit{Trends promoted more by one age group}\\
The last three rows in Table~\ref{tab:niche_interest} show the trends promoted by either mostly Adolescents, Young, or mostly Mid-aged people.
 Examples of topics promoted by adolescents include
\#WhoSaysYouAreNotPerfectSelena, which is about the celebrity Selena Gomez, or \#NationalTeddyBearDay.
Trends promoted predominantly by young people include  
health related issues like \#breastfeeding, or \#WWEChicago (a wrestling match). Finally,
the trends promoted by mid-aged people, tend to cover
many political topics such as \#HillaryHealth, \#TrumpInDetroit, and \#WheresYourTaxes.

\vspace{1mm}
\noindent \textbf{Trends expressing different perspectives during \\ different events} \\
Apart from the niche interests, trends promoted by different demographic groups 
also tend to offer unique perspectives during events relevant to the broad society. 
Here, we discuss the topics which became trending in Twitter during the following three events: \\
A. US Independence Day (on 4th July, 2016) \\
B. Dallas Shooting (on 7th July, 2016) \\
C. US Presidential Election 2016 (on 8th November, 2016) \\

\noindent \textit{US Independence Day} \\
July Fourth is the independence day of the US, and therefore, many related topics became trending. However, the long election campaigns as well as recent increase in racial tensions have prompted different Twitter users to promote different trends expressing their views. Table~\ref{tab:independence_day} shows the demographics of promoters of different associated trends. We can see that trends like \#AmericaWasNeverGreat was promoted by mid-aged black people, similarly, \#WeAreAmerica was promoted by young black men.
\#GrowingUpInABlackChurch was promoted by adolescent black women. On the other hand, trends such as \#Freedom, \#GodBlessAmerica were promoted by young white men. \\

%\vspace{1mm}
\noindent \textit{Dallas Shooting} \\
On 7th July, 2016, a protest was organized in Dallas, Texas by the group `Black Lives Matter', against the killings of two black men, Alton Sterling and Philando Castile, by police officers in Louisiana and Minnesota, few days before. During the protest, 5 police officers in Dallas were assassinated by a black army veteran Micah Xavier Johnson\footnote{nytimes.com/2016/07/08/us/dallas-police-officers-killed.html}. Subsequently, on 8th July, police killed Johnson with a remote controlled bomb carried by a robot. This event also marked the first use of a robot to kill a suspect by police in US\footnote{talkingpointsmemo.com/livewire/suspect-killed-bomb-robot}.

In Table~\ref{tab:dallas_shooting}, we show trends which were promoted by users having different demographics. We can clearly see how different trends express different perspectives on the same event. \#DallasPoliceShootings, and \#PoliceLivesMatter were promoted by young or mid-aged white users. \#BattleBots was promoted by young and mid-aged men across all races. On the other hand, \#BlackLivesMatter, and  \#BlackSkinIsNotACrime were promoted by adolescent and young black people. \#AllLivesShouldMatter was promoted by a combination of black and asian adolescents. Finally, \#PrayForPeace was promoted by young white women.\\

\noindent \textit{US Presidential Election 2016} \\
US presidential election of 2016 was held on 8th November, 2016, where the major contenders were Democratic candidate Hillary Clinton, and Republican candidate Donald Trump. The election results became clear on 9th November early morning, with Trump becoming the president-elect. In addition to the dataset described in the dataset section, we collected trends, tweets and the demographic information of Twitter users participating in the trends during the election period. In Table~\ref{tab:us_election}, we present the election related trends chronologically.

We can see that on 7th November, before the election, the election related trends were promoted by mostly young and mid-aged white people. The political distinctions can be seen in the gender of the promoters. While \#ImWithHer, Clinton campaign slogan, was promoted by both men and women; promoters of \#TrumpWinsBecause was dominated by men. On the election day, the trends were mostly promoted by young people, and by white men, white women and black men.

On the day of the election result, we see multiple trends emerging, each representing the perspectives of different groups. For example, \#ElectionResults, \#PresidentTrump and \#TrumpsFirstOrder were promoted by mid-aged white men. On the other hand, \#RIPAmerica, and \#ImStillWithHer were promoted by adolescent and young white women.
Finally, \#NowThatTrumpIsPresident was promoted by young and mid-aged black and asian men.

\section{Concluding Discussion}
\noindent In this paper, we focused on understanding the demographics
of crowds who make some content worthy of being recommended as
trending.  We particularly focus on the promoters of Twitter
trends. Using an extensive dataset from Twitter, we analyzed how the
promoters of different trends compare with the overall Twitter
population. 

Our analysis shows that a large fraction of Twitter trends are promoted
by users, whose demographic composition differs significantly from
Twitter's user population. More troublingly, we find that
traditionally marginalized social groups (e.g., black women) are
systematically under-represented among the promoters of Twitter
trends. We observe that the trends predominantly promoted by a specific
demographic group either tend to be of niche interest or reflect
divergent perspectives on events of broad societal interest.

Our work adds an important perspective to ongoing debates about the
fairness, bias, and transparency of algorithms operating over big
crowdsourced datasets: {\it their inputs}. Our findings show that
beyond studying algorithms and their outputs (e.g., search results,
trending topics), it is useful to understand the inputs over which the
algorithms work (e.g., characteristics of the crowd who make
a topic popular).

\vspace{2mm}
\noindent \textbf{Making demographic biases in trends transparent} \\   
Given our findings that (i) the demographics of the crowds promoting
trends is often quite different from the overall population of the
media site, and (ii) the demographics of promoters has a strong influence
on what topics will become trending, there is a clear need to make the
demographic biases in trend recommendations transparent to Twitter
users.

Towards that end, we developed and publicly deployed a system
`Who-Makes-Trends'\footnote{\tt twitter-app.mpi-sws.org/who-makes-trends} 
to make the demographics
of crowds promoting Twitter trends in the US more transparent.
We believe that such systems are not only useful for the social media
users, but also for journalists, social media researchers, developers
of recommendation systems, as well as for governmental agencies
wanting to understand different facets of public opinion during
moments of unrest.

\vspace{1mm}
\noindent \textbf{Directions for future work} \\ 
While our study here is limited to understanding biases in the inputs to the crowdsourced
recommendations (i.e., trending topics), we believe that our analysis
framework can be easily extended and our core findings will be
relevant to a variety of algorithms in social media that rely on
inputs from crowds, including social search~\cite{Kulshrestha2017} and
assessing reputation or influence of users in social media. Another
avenue for future work lies in investigating new algorithms for
selecting trending topics that explicitly take into account the
demographic biases of the crowds promoting individual topics.

\vspace{0.5cm}
\noindent\textbf{Acknowledgments:}
This research was partly supported by a grant from the IMPECS project
titled {\it Understanding, Leveraging and Deploying Online Social Networks}.
F. Benevenuto was supported by grants from Humboldt Foundation, Capes, CNPq, and Fapemig.
A. Chakraborty is a recipient of Google India PhD Fellowship and
Prime Minister's Fellowship Scheme for Doctoral Research,
a public-private partnership between Science \& Engineering Research Board (SERB),
Department of Science \& Technology, Government of India and
Confederation of Indian Industry (CII).

{\small
\bibliographystyle{aaai}
\bibliography{Main}
}

\end{document}